\documentclass[twocolumn,showpacs,amsfonts,aps,prc,nofootinbib,floatfix,%
superscriptaddress]{revtex4}

\usepackage{amsmath}
\usepackage{bm}
\usepackage{graphicx}

\usepackage{epsfig}
\newcommand{\beq}{\begin{equation}}
\newcommand{\eeq}{\end{equation}}
\newcommand{\bea}{\vspace{0.25cm}\begin{eqnarray}}
\newcommand{\eea}{\end{eqnarray}}

\newcommand{\ro}{\mbox{{\boldmath
$\rho$}}}

\newcommand{\pb}{\mbox{{\bf
p}}}

\newcommand{\kb}{\mbox{{\bf
k}}}

\newcommand{\qbt}{\mbox{{\bf
q}}_\perp}

\def\lsim{\mathrel{\rlap{\lower4pt\hbox{\hskip1pt$\sim$}}
    \raise1pt\hbox{$<$}}}         
\def\gsim{\mathrel{\rlap{\lower4pt\hbox{\hskip1pt$\sim$}}
    \raise1pt\hbox{$>$}}}         

\newcommand{\landau}{L.D.~Landau Institute for Theoretical Physics,
        GSP-1, 117940, Kosygina Str. 2, 117334 Moscow, Russia}

\begin{document}


\title{
Phenomenology of collinear photon emission from quark-gluon plasma
in $AA$ collisions
}
\date{\today}

\author{B.G.~Zakharov}\affiliation{\landau}

\begin{abstract}
We study the role of running coupling and 
the effect of variation of the thermal quark mass on
contribution of the collinear bremsstrahlung and annihilation 
to photon emission in $AA$ collisions in a scheme 
similar to that used in our previous jet quenching analyses.
We find that 
for a scenario with the thermal quark mass
$m_q\sim 50-100$ MeV  contribution of the
higher order collinear processes summed with the $2\to 2$ processes
can explain a considerable part ($\sim 50$\%) 
of the experimental photon spectrum 
at $k_T\sim 2-3$ GeV
for Au+Au collisions at $\sqrt{s}=0.2$ TeV.
But  for $m_q=300$ MeV and for the 
thermal quark mass predicted by the HTL scheme 
the theoretical predictions
underestimate 
considerably the experimental spectrum.

\end{abstract}
%
\maketitle

\section{Introduction}

The observation 
of jet quenching phenomenon
and hydrodynamical flow effects
in $AA$ collisions at RHIC and LHC
signals about formation of a hot quark-gluon plasma (QGP)
in the initial stage of $AA$ collisions. It seems likely
that the QGP formation goes via the thermalization of the 
collective color fields of the so-called glasma stage
\cite{glasma1,glasma2} formed after multiple gluon exchanges between 
two strongly Lorentz contracted nucleus disks.  
It is believed that the QGP 
should also reveal itself in thermal photon emission
that may be important in the low and 
intermediate $k_T$ region \cite{Shuryak}.
However, the 
photon production in $AA$ collisions 
shows some inconsistency with 
the QGP evolution supported by 
the results of the jet quenching 
analyses. 
The data from RHIC and LHC on jet quenching in $AA$ collisions  
can be explained in the picture
with radiative and collisional energy loss  
for the hydrodynamical 
QGP evolution with the QGP production time $\tau_0\sim 0.5$ fm
and the initial entropy determined from the measured 
hadron multiplicities  \cite{RAA12,UW_JQMC,CUJET}. However,
theoretical predictions for the thermal photon spectrum in this picture 
obtained with a sophisticated viscous hydrodynamical model of 
the fireball evolution \cite{Gale_best}  underestimate the photon 
spectrum measured at RHIC by PHENIX \cite{PHENIX_ph_PR} 
in Au+Au collisions at $\sqrt{s}=0.2$ TeV   
by a factor of $\sim 3$. Several mechanisms have been suggested 
that can increase the photon emission in $AA$ collisions.
There were suggestions that very strong
magnetic field created in noncentral $AA$ collisions 
can increase the
photon emission due to the conformal anomaly \cite{Kharzeev1} and
the synchrotron radiation \cite{T1}. However, these mechanisms
require too high magnitude of the magnetic field \cite{Z_syn2}, that contradicts
to calculations for realistic evolution of the plasma fireball \cite{Z_maxw}.
In Ref. \cite{Snigirev} it was suggested that a considerable additional
contribution to the photon
production may be due to the boundary bremsstrahlung 
resulting from interaction of escaping quarks with collective
confining color field at the surface of the QGP. 
In Refs. \cite{glasma_L,glasma_L2,glasma_W} it was 
argued that the pre-equilibrium glasma phase also can give 
large contribution to the photon emission in $AA$ collisions.  
Unfortunately, uncertainties in the theoretical predictions 
for the boundary photon emission \cite{Snigirev} and the photon
emission from the glasma \cite{glasma_L,glasma_L2,glasma_W} are rather large.

As compared to the glasma stage the photon production in the QGP 
stage is better understood. However, even for the QGP phase the
theoretical uncertainties can be considerable, because
the available analyses are based on the pQCD picture of a weakly coupled QGP.
And its applicability to the QGP produced at RHIC and LHC may be questionable.  
In the leading order (LO) pQCD the thermal photon emission from the QGP 
 is due to the
$2\to 2$ processes: $q(\bar{q})g\to \gamma q(\bar{q})$ (Compton)
and $q\bar{q}\to \gamma g$ (annihilation). 
In the pQCD picture a significant contribution to the photon 
emission comes also 
from the higher order collinear processes $q\to \gamma q$ and $q\bar{q}\to
\gamma$ \cite{AMY1}. It turns out to be parametrically of the same
order as the $2\to 2$ processes \cite{AGZ2000}. The results of
Ref. \cite{AMY1}
show that at $k/T\gsim 3$ contribution
of the collinear processes turns out to be close to that from 
the LO $2\to 2$ mechanisms, and  at $k/T\lsim 2$ 
the collinear emission gives the dominant contribution to the photon emission
rate in the QGP.
The collinear photon radiation is due to 
multiple scattering of thermal quarks in the QGP. This mechanism 
is similar to that for the photon radiation 
from hard quarks \cite{Z_phot}
and to the induced gluon radiation from fast partons that
dominates in the jet quenching phenomenon 
\cite{GW,BDMPS,LCPI1,Z_1998rev}. 
In \cite{AMY1} the collinear processes have been evaluated 
for constant QCD coupling using the 
thermal field theory methods within the hard thermal loop (HTL)
resummation scheme. 
In the case of the
induced gluon emission from fast partons in the QGP
the results for constant and running $\alpha_s$ differ considerably.  
For running $\alpha_s$ the energy dependence of the 
radiative parton energy loss weakens \cite{Z_Ecoll}. The analyses of the
data on the nuclear modification factor $R_{AA}$ from RHIC and LHC
\cite{Z_Ecoll,RAA08,CUJET} show that running $\alpha_s$ allows to
obtain a better agreement with the data. In \cite{Gale_best} the photon 
emission has been addressed using the AMY \cite{AMY1} formulas
obtained for a fixed QCD coupling constant.
For accurate confronting the QGP signals from jet quenching and from photon
production it would be of great interest to perform calculations 
of the collinear photon emission with running $\alpha_s$ consistent with
that used in the successful jet quenching analyses. 
Also it would be interesting to study the sensitivity of the collinear
photon emission to variation of the quark quasiparticle mass $m_q$. 
The predictions of the pQCD analysis \cite{AMY1}, based on the HTL 
resummation scheme,  have been obtained for the standard pQCD 
quark quasiparticle mass $m_q=gT/\sqrt{3}$. However, the analysis  of
the lattice data within a quasiparticle model \cite{LH}
gives practically constant thermal quark mass $m_q\sim 300$ MeV. 
In a more recent analysis 
\cite{mq_sQGP} it was demonstrated that in a strongly coupled QGP
the thermal quark mass may be much smaller than that in the pQCD 
HTL picture. A two-pole fit (with the normal and plasmino modes) 
of the Euclidean lattice quark correlator
of Ref. \cite{Karsch_mq} also supports that the thermal quark mass
may be smaller than in the HTL scheme (by a factor of $\sim 2$). However,
unfortunately the fit is not very accurate due to 
the insensitivity of the Euclidean correlator to the quark
spectral function at energies $\lsim T$ \cite{Karsch_mq}.    
The small thermal quark mass may increase the photon emission rate,
with a very small effect on the jet quenching that is practically insensitive
to the quark quasiparticle mass \cite{BDMPS,LCPI1}. 
Due to the theoretical uncertainties for the thermal quark mass, 
it would be interesting to study the collinear photon
emission in a phenomenological picture without the HTL constraints 
on the quark quasiparticle mass.    

In the present paper we study the effect of running $\alpha_s$ 
and the role of variation of the quark quasiparticle mass on 
the collinear photon emission in $AA$ collisions.
We treat quark multiple scattering in the QGP in the scheme we used previously
in successful jet quenching analyses \cite{RAA08,RAA12}.
There we used the Debye mass obtained in the lattice calculations
that, contrary to the HTL scheme, give nonzero magnetic screening
\cite{magnetic_Md} in the QGP.
We compare the results for this scenario with the  
results for the HTL scheme with static $\alpha_s$.
We use the formalism of \cite{AZ}
based on the light-cone path integral (LCPI) approach
\cite{LCPI1,Z_1998rev}. The formulation given in \cite{AZ} reproduces
the results of the AMY \cite{AMY1} approach.
In \cite{AMY1} the photon emission rate has been expressed via
solution of an integral equation. In the present paper 
the photon emission rate is expressed via solution
of a two-dimensional Schr\"odinger equation with a smooth boundary
condition. The method is convenient for numerical calculations.

\section{Theoretical framework}
The contribution of 
the collinear processes $q\to \gamma q$ and $q\bar{q}\to \gamma$
to the photon emission 
rate per unit time and volume in the plasma rest frame 
can be written as \cite{AZ,AMY1}
\beq
\frac{dN}{dtdVd\kb}=
\frac{dN_{br}}{dtdVd\kb}
+\frac{dN_{an}}{dtdVd\kb}\,,
\label{eq:10}
\eeq
where the first term corresponds to $q\to \gamma q$ and the second one
to $q\bar{q}\to \gamma$.
The bremsstrahlung contribution can be written as \cite{AZ}
\bea
\frac{dN_{br}}{dtdVd\kb}=\frac{d_{br}}{k^{2}(2\pi)^{3}}
\sum_{s}
\int_{0}^{\infty} dp p^{2}n_{F}(p)
\nonumber\\ \times
[1-n_{F}(p-k)]\theta(p-k)
\frac{dP^{s}_{q\rightarrow \gamma q}(\pb,\kb)}{dk dL}\,,
\label{eq:20}
\eea
where 
$d_{br}=4N_{c}$ is the number of the quark and antiquark states,
\beq
n_{F}(p)=\frac{1}{\exp(p/T)+1}\,
\label{eq:30}
\eeq
is the thermal Fermi distribution, 
and 
${dP^{s}_{q\rightarrow \gamma q}(\pb,\kb)}/{dk dL}$
is the photon emission probability distribution 
per unit length for a quark of type $s$.
In the small angle approximation 
we can take the vectors $\pb$  and $\kb$ parallel.
The annihilation contribution 
can be expressed via the probability
distribution for the photon absorption
with the help of the detailed balance principle.
It leads to the formula \cite{AZ} 
\bea
\frac{dN_{an}}{dtdVd\kb}=\frac{d_{an} }{(2\pi)^{3}}
\sum_{s}
\int_{0}^{\infty} dp n_{F}(p)
\nonumber\\ \times
n_{F}(k-p)\theta(k-p)
\frac{dP^{s}_{\gamma\rightarrow q\bar{q}}(\kb,\pb)}{dp dL}\,,
\label{eq:40}
\eea
where $d_{an}=2$ is the number
of the photon helicities, 
$
{dP^{s}_{\gamma\rightarrow q\bar{q}}(\kb,\pb)}/{dp dL}
$ is the probability distribution per unit length
for the $\gamma \rightarrow q\bar{q}$ transition
($p$ is the quark momentum and $k-p$ is the antiquark momentum,
and similarly to $q\to \gamma q$ we can take the vectors 
$\pb$  and $\kb$ parallel).

In the LCPI formalism \cite{LCPI1} 
the probability of the $q\to \gamma q$ transition
per unit length (in terms of the fractional photon momentum $x=k/p$)
can be written as
\bea
\frac{d P_{q\rightarrow \gamma q}}{d
x dL}=2\mbox{Re}
\int\limits_{0}^{\infty} d
z
\exp{\left(-i\frac{z}{L_{f}}\right)}
\hat{g}(x)
\nonumber\\ \times
\left[
{\cal K}(\ro_{2},z|\ro_{1},0)
-{\cal K}_{vac}(\ro_{2},z|\ro_{1},0)
\right]\bigg|_{\ro_{1,2}=0}\,,
\label{eq:50}
\eea
where $L_{f}=2M(x)/\epsilon^{2}$ with $M(x)=E_qx(1-x)$,
$\epsilon^{2}=m_{q}^{2}x^{2}+m_{\gamma}^{2}(1-x)$
(in general for $a\to b+c$ transition 
$\epsilon^{2}=m_{b}^{2}x_{c}+m_{c}^{2}x_{b}-m_{a}^{2}x_{b}x_{c}$),
$\hat{g}$ is the vertex operator, given by
\beq
\hat{g}(x)=\frac{V(x)}{M^{2}(x)}\frac{\partial }{\partial \ro_{1}}\cdot
\frac{\partial }{\partial \ro_{2}}\,
\label{eq:60}
\eeq
with 
\beq
V(x)=z_{q}^{2}\alpha_{em}(1-x+x^{2}/2)/x, 
\label{eq:70}
\eeq
$\alpha_{em}=e^2/4\pi$ the fine-structure constant.  In (\ref{eq:50}) 
$\cal{K}$ is the retarded Green's function of a two dimensional
Schr\"odinger equation
with the Hamiltonian
\beq
\hat{\cal{H}}=-\frac{1}{2M(x)}
\left(\frac{\partial}{\partial \ro}\right)^{2}
+         v(\ro)\,,
\label{eq:80}
\eeq
and 
${\cal{K}}_{vac}$ is the Green function for $v=0$.
The potential $v$ can be written as
\beq
v=-i P(x\rho)\,,
\label{eq:90}
\eeq
where the function $P(\rho)$ describes interaction of the color singlet 
$q\bar{q}$ dipole with the QGP. In the HTL scheme with static coupling
constant $g$ 
$P(\rho)$ can be written as \cite{AZ,PA_C}
\beq
P(|\ro|)= \frac{g^{2}C_{F}T}{(2\pi)^{2}}\int d\qbt [1-\exp(i\ro \qbt)]
C(\qbt)\,,
\label{eq:100}
\eeq
\beq
C(\qbt)=\frac{m_{D}^{2}}{\qbt^{2}(\qbt^{2}+m_{D}^{2})}\,,
\label{eq:110}
\eeq  
where $C_F=4/3$ is the quark Casimir, $m_{D}=gT[(N_{c}+N_{F}/2)/3]^{1/2}$ 
is the Debye mass. 
In the approximation of static color Debye-screened scattering centers 
\cite{GW} the function $P(\rho)$ reads
\beq
P(\rho)= 
\frac{n{\sigma}(\rho )}{2}\,,
\label{eq:120}
\eeq
where $n$ is the number density
of the color centers, and $\sigma(\rho )$
is the well known dipole cross section. For running
$\alpha_s$ the dipole cross section reads \cite{NZ12}
\beq
\sigma(|\ro|)={C_{T}C_{F}}\int d\qbt \alpha_{s}^{2}(q_{T}^2)
\frac{[1-\exp(i\qbt\ro)]}{(\qbt^{2}+m_{D}^{2})^{2}}\,,
\label{eq:130}
\eeq
where $C_T$ is the color center Casimir.
The dipole form (\ref{eq:120}) was used in
our previous jet quenching analyses \cite{RAA08,RAA12} with $\alpha_s(q^2)$ 
frozen at some value $\alpha_{s}^{fr}$ at low momenta.

For numerical calculations it is convenient to use the representation
of the spectrum 
as a sum of the Bethe-Heitler term and an absorptive correction
due to higher order rescatterings (describing the Landau-Pomeranchuk-Migdal
suppression \cite{Z_1998rev})    
\beq
\frac{d P_{q\to \gamma q}}{d
xdL}=
\frac{d P_{q\to \gamma q}^{BH}}{d x dL}+
\frac{d P_{q\to \gamma q}^{abs}}{d
x dL}\,.
\label{eq:140}
\eeq
It can be derived 
by expanding the Green's function ${\cal K}$ in
(\ref{eq:50})
in a series in the potential
$v$ (see \cite{Z_SLAC1} for details).
The Bethe-Heitler contribution corresponds to the term linear in $v$.
It can be written as 
\beq
\frac{d P_{q\to \gamma q}^{BH}}{d
xdL}=\frac{n}{2}
\sum\limits_{\{\lambda\}} \int
d\ro\,
|\Psi(x,\ro,\{\lambda\})|^{2}
\sigma(\rho
x)\,\,,
\label{eq:150}
\eeq
where $\{\lambda\}=(\lambda_{q},\lambda_{q'},
\lambda_{\gamma})$ is the set of helicities,
$
\Psi(x,\ro,\{\lambda\})
$
is the light-cone wave function for $q\to \gamma q$ transition 
with $\lambda_{q'}=\lambda_{q}$ ).
The contribution of the higher order rescatterings reads
\bea
\frac{d P_{q\to \gamma q}^{abs}}{d
x dL}=-\frac{n^2}{4}\mbox{Re}
\sum\limits_{\{\lambda\}}
\int\limits_{0}^{\infty}
dz
\int
d\ro\,
\Psi^{*}(x,\ro,\{\lambda\})\nonumber\\
\times\sigma(\rho
x)
\Phi(x,\ro,\{\lambda\},z,0)
\exp\left(-\frac{iz}{L_{f}}\right)\,,
\label{eq:160}
\eea
where
\bea
\Phi(x,\ro,\{\lambda\},z_{2},z_1)=
\int d\ro'{\cal
K}(\ro,z_{2}|\ro',z_{1})
\nonumber\\ \times
\Psi(x,\ro',\{\lambda\})\,\sigma(\rho'
x)
\label{eq:170}
\eea
is the solution of the Schr\"odinger equation with the
boundary
condition
$
\Phi(x,\ro,\{\lambda\},z_{1},z_{1})=
\Psi(x,\ro,\{\lambda\})\sigma(\rho
x)\,.
$

For $\gamma\to q \bar{q}$ one can obtain similar formulas.
But now $M(x)=E_{\gamma}x(1-x)$ ($x$ is the quark fractional momentum)
$\epsilon^{2}=m_{q}^{2}-m_{\gamma}^{2}x(1-x)$, and 
\beq
V(x)=z_{q}^{2}\alpha_{em}N_c[x^{2}+(1-x)^{2}]/2\,.
\label{eq:180}
\eeq
The formulas for the light-cone wave functions for the 
$q\to \gamma q$ and $\gamma\to q\bar{q}$ are similar to that
that for the QED processes $e\to \gamma e$ and $\gamma\to e\bar{e}$
given in \cite{Z_1998rev}.

We will present the results for two versions 
of the model: for the phenomenological scenario with running $\alpha_s$
and for the pQCD HTL scenario with static 
coupling \cite{AMY1}. 
For the scenario with running coupling, as in our jet quenching analyses,
we use the dipole formulas (\ref{eq:120}), (\ref{eq:130}). 
In jet quenching analyses \cite{RAA08,RAA12} we used  
the quark quasiparticle mass $m_{q}=300$ MeV. 
For the relevant temperature region $T\sim (1-3)T_c$,
it is supported by the analysis of Ref. \cite{LH}  of the lattice data in 
the quasiparticle model. 
For the induced gluon emission the results are practically
insensitive to the light quark mass. But for the photon emission
the value of the quark mass is important.  
As was shown recently in Ref.~\cite{mq_sQGP}, in a strongly coupled 
QGP the thermal quark mass may be much smaller 
than the pQCD prediction based on the HTL scheme. 
Therefore, for the phenomenological scenario we present the results for two 
very different values of the thermal quark mass: $m_{q}=300$ MeV (as 
obtained in Ref.~\cite{LH}) and $m_q=50$ MeV. The latter value is much smaller
than the thermal pQCD HTL quark mass, and the results should be close to 
that for the massless quarks supported by the analysis \cite{mq_sQGP}. 
As in jet quenching analyses, for the version with running $\alpha_s$ we use 
the Debye mass obtained in the lattice calculations \cite{Bielefeld_Md}, that 
give $m_{D}/T$ slowly decreasing with $T$  
($m_{D}/T\approx 3$ at $T\sim 1.5T_{c}$, $m_{D}/T\approx 2.4$ at 
$T\sim 4T_{c}$). 
For the pQCD HTL scenario we use for the quark and Debye masses the standard
pQCD values ($m_q=gT/\sqrt{3}$, $m_{D}=gT[(N_{c}+N_{F}/2)/3]^{1/2}$),
and the formulas (\ref{eq:100}), (\ref{eq:110})  for the function $P(\rho)$.
To account for the mass suppression for strange quarks we take $N_f=2.5$
as in our jet quenching analyses \cite{RAA12}.

\section{Numerical results for photon spectrum in $AA$ collisions}
In calculating the photon spectrum in $AA$ collisions
we perform the four volume integration 
using the proper time $\tau$ and rapidity $Y$ variables 
\beq
\tau=\sqrt{t^2-z^2}\,,\,\,\,\, 
Y=\frac{1}{2}\ln\left(\frac{t+z}{t-z}\right)\,.
\label{eq:190}
\eeq
In these coordinates the photon spectrum reads 
\beq
\frac{dN}{dy d\kb_T}=\int \tau d\tau dY d\ro\, \omega'
\frac{dN(T',k')}{dt'dV'd\kb'}\,,
\label{eq:200}
\eeq
where primed quantities correspond to the comoving frame, and 
$\omega'=k'=|\kb'|$.

We describe the plasma fireball 
at $\tau>\tau_0$ 
in the Bjorken model \cite{Bjorken} without the transverse expansion
that gives the entropy density $s\propto 1/\tau$.
We present the results for the ideal gas model (with 
the temperature dependence of the entropy density $s\propto T^3$), that gives  
$
T=T_0(\tau_0/\tau)^{1/3}
$ in the plasma phase. We also perform calculations for the temperature
dependence of the entropy density $s(T)$ obtained in the lattice simulation
\cite{EoS}.
As in jet quenching analyses \cite{RAA12} we take $\tau_0=0.5$ fm.
To account for qualitatively the fact that 
the process of the QGP production is not instantaneous, 
we take the entropy density $\propto \tau$ in the interval $0<\tau<\tau_0$.
However, the contribution of this region is relatively small.
We calculate the initial 
density profile of the QGP fireball 
at the proper time $\tau_{0}$
assuming that the initial entropy is proportional to 
the charged particle pseudorapidity 
multiplicity density at $\eta=0$ 
calculated in the two component wounded nucleon 
Glauber model \cite{KN} 
(the details and model parameters can be found in \cite{Z_syn2,Z_MCGL}).
In the space-time integral (\ref{eq:200}) we drop the points 
with $T_0<T_c$ (here $T_c=160$ MeV is the deconfinement temperature) 
at $\tau=\tau_0$. 
For the ideal gas model we treat the crossover region at 
$T\sim T_c$ as a mixed phase,
and take the entropy density in this phase $\propto 1/\tau$ \cite{Bjorken}.
In the mixed phase we account for the photon emission only from 
the QGP phase. Note that contribution of the space-time region with $T\sim T_c$ 
to the photon spectrum 
(both for the ideal gas fireball model and for the lattice version of 
the entropy 
density) is relatively small 
at $k_T\gsim 1.5-2$ GeV.

\begin{figure} 
\begin{center}
\epsfig{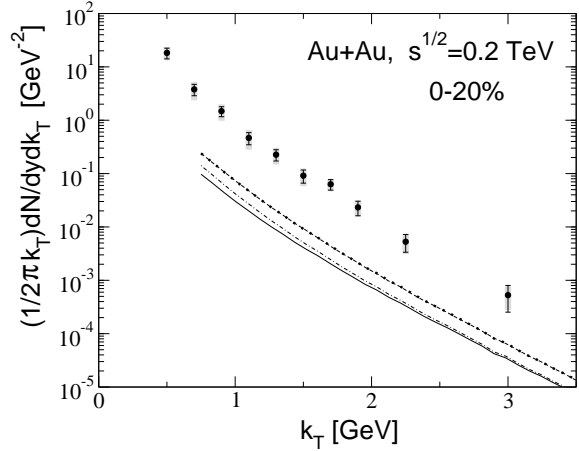}
\end{center}
\caption[.]{
The photon spectrum  $(1/2\pi k_T)dN/dydk_T$ 
averaged over the azimuthal angle for Au+Au collisions at 
$\sqrt{s}=0.2$ TeV in the $0-20$\% centrality range.
Solid: the sum of the $q\to \gamma q$ and $q\bar{q}\to \gamma$
processes for running coupling with $\alpha_s^{fr}=0.5$ 
for $m_q=300$ MeV,
dotted:  the same as solid but for $m_q=50$ MeV,
dot-dashed: the sum of the $q\to \gamma q$ and $q\bar{q}\to \gamma$
processes for the HTL scheme for $\alpha_s=0.3$,
dashed: the sum of the collinear process with the LO $2\to 2$ 
processes for the HTL scheme for $\alpha_s=0.3$.
The theoretical curves are for the ideal gas model for $\tau_0=0.5$ fm. 
Data points are from Ref.~\cite{PHENIX_ph_PR}.
}
\end{figure}
\begin{figure} 
\begin{center}
\epsfig{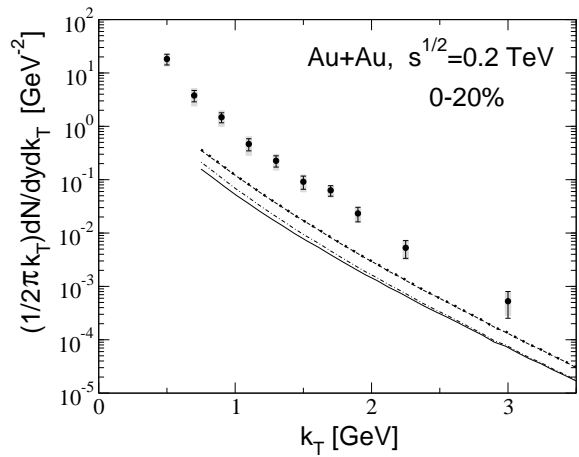}
\end{center}
\caption[.]{
Same as in Fig.~1 but for calculations with the entropy density $s(T)$ from
the lattice simulation \cite{EoS}.
}
\end{figure}

For the phenomenological scenario with running QCD coupling 
we assume that at low momenta $\alpha_s$ is 
frozen at the value $\alpha_{s}^{fr}=0.5$,
that is supported by our jet quenching analyses of the 
nuclear modification factors $R_{AA}$ \cite{RAA12} 
and $I_{AA}$ \cite{Z_IAA} 
in Au+Au collisions at 
$\sqrt{s}=0.2$ TeV.
For the HTL scenario with fixed coupling we take $\alpha_s=0.3$. 
In Fig.~1 we show our results for the ideal gas model of the QGP
for the photon
spectrum $dN/dy d\kb_T=(1/2\pi k_T)dN/dy dk_T$ 
(averaged over the azimuthal angle)
for Au+Au collisions at $\sqrt{s}=0.2$ TeV for 
$0-20$\% centrality bin for the phenomenological scenario
with running coupling for $m_q=300$ and $50$ MeV and for the HTL scenario.
We compare our results with the data from PHENIX 
\cite{PHENIX_ph_PR}.
The theoretical curves have been obtained integrating 
in (\ref{eq:200}) 
up to 
$\tau_{max}=10$ fm. 
At $k_T\gsim 1$ GeV the photon spectrum is only weakly sensitive 
to $\tau_{max}$. It occurs because the main contribution at $k_T\gg T_0$ comes 
from the hottest space-time region of the QGP with $\tau$ up to
several units of $\tau_0$.
For $\tau_{max}=R_A\approx 6.4$ fm the photon spectrum
at $k_T\sim 1$ GeV  is reduced only by $\sim 10$\% 
and for $k_T\gsim 2$ GeV the change in the spectrum is negligible.    
For the HTL scenario we also present in Fig.~1 the sum of the 
contributions from the collinear 
processes and from  the LO contribution due to $2\to 2$ processes 
in the form obtained in \cite{AMY1}.
From Fig.~1 one can see that 
the results for the phenomenological scenario with running coupling 
and $m_q=300$ MeV are close to that for the HTL scenario with fixed coupling.
But for the phenomenological scenario with $m_q=50$ MeV the photon
spectrum is bigger than that for the HTL scenario by a factor of $\sim 2$.
Note that the photon yields obtained for $m_q=300$ and 
$50$ MeV do not follow the power low $1/m_q^2$. This is due to
the Landau-Pomeranchuk-Migdal suppression, 
described by the absorptive term on the right-hand side of (\ref{eq:140}),
that becomes very strong for small $m_q$. In this regime the quark mass
dependence becomes very weak. Our calculations show that the photon spectrum
for $m_q=100$ MeV is smaller than that for $m_q=50$ MeV only by $\sim 20$\%. 
From Fig.~1 one can see that at $k_T\gsim 1.5-2$ 
GeV for the HTL scenario  the theoretical curves 
for the sum of the contribution from the collinear processes
$q\to \gamma q$ and $q\bar{q}\gamma$ and the LO mechanisms
underpredict the data typically by a factor of $\sim 5-7$.
Assuming that for the phenomenological scenario the relative 
effect of the $2\to 2$ processes is similar to that for the HTL scenario
\footnote{The incorporation of running $\alpha_s$ for 
the $2\to 2$ processes 
was not elaborated yet. However, calculations
in the HTL scheme with static $\alpha_s$ show that contribution
of the LO processes has relatively low sensitivity to $\alpha_s$
(say, for $\alpha_s=0.3$ the growth of the LO contribution 
as compared to that for $\alpha_s=0.2$ is $\lsim 10-25$\%). 
Therefore, the effect of running coupling constant 
on the $2\to 2$ processes should not be very large.
Note that even for the scenario with a very small thermal  quark  mass 
\cite{mq_sQGP} the contribution of the $2\to 2$ processes should not 
change significantly. Because it depends logarithmically on the 
quark quasiparticle mass. And even for a vanishing quasiparticle mass 
in an infinite QGP, for the $2\to 2$ process in the expanding QGP 
the effective
quark virtuality cannot be much smaller than $1/\tau_{ev}$, where 
$\tau_{ev}\sim 1-4$ fm is the typical QGP evolution time
dominating the photon emission.
},
we can conclude that even for the version with $m_q=50$ MeV
the experimental spectrum will be underestimated by a factor of $\sim 3$.
The situation becomes better for the results obtained for 
the lattice temperature dependence of the entropy density, that are shown
in Fig.~2. In this case the theoretical predictions are approximately 
increased by a factor of $\sim 1.5-2$, and the disagreement with the data
becomes smaller. The inclusion of the hadron gas phase \cite{Gale_HG}
can improve the agreement with the data at low $k_T$ ($k_T\lsim 1$
GeV). But it cannot increase significantly the photon spectrum 
at $k_T\sim 2-3$ GeV. Thus, we can conclude that  
in this high-$k_T$ region, even for the scenario with a small thermal  
quark  mass, 
one cannot avoid  some underestimation of the photon spectrum.  

The agreement with the data at high $k_T$ can be improved
for a smaller value of the thermalization time $\tau_0$.
Our calculations for $\tau_0=0.25$ fm show that the theoretical predictions
increase by a factor of $\sim 2$ at $k_T\sim 2-3$ GeV.
However, such a small value of $\tau_0$ does not seem realistic,
because it is of the order of the inverse saturation scale $1/Q_s$
($Q_s\sim 1-1.5$ GeV for RHIC conditions \cite{Lappi_qs}).
And in this time region the description in terms of the 
pre-equilibrium glasma stage is more
appropriate. The considerable  increase of the photon spectrum 
for $\tau_0=0.25$ fm  can be viewed as an indication that 
the glasma phase contribution to the photon production at $k_T\gsim 2$ GeV 
can be large. In fact, the glasma effect can be considerably bigger. 
Because the typical Lorentz force that quarks undergo
in the glasma is by a factor of $\sim 10-20$ bigger than that 
for the Debye screened color centers in the thermalized QGP \cite{AZ_glasma}.
However, due to finite formation length of the collinear photon emission
an accurate analysis of the collinear processes including the 
pre-equilibrium glasma stage is a complicated task. Because, 
due to the nonlocal nature the photon emission, the photon spectrum 
should be sensitive
to the whole process of the QGP formation, and one simply cannot distinguish
the photon emission from the glasma and from the QGP at $\tau\sim \tau_0$.  
It worth to note that the magnitude of the jet quenching is not
strongly affected by the variation of $\tau_0$ from $\sim 0.5$ fm 
to $\sim 0.25$. It is due to a strong reduction of the 
radiative parton energy loss by the finite size effects for the first fm/c of
the matter evolution \cite{Z_OA}. For the same reason the glasma effect
on jet quenching also turns out to be small \cite{AZ_glasma}.

\section{Summary}
We have studied the role of running coupling and 
the effect of variation of the thermal quark mass on
contribution of the collinear processes
$q\to \gamma q$ and $q\bar{q}\to \gamma$ in the QGP phase to the photon
spectrum in $AA$ collisions in a phenomenological scheme 
similar to that used in our previous successful jet quenching analyses
based on the LCPI approach \cite{LCPI1} to the induced gluon emission.
The analysis of the collinear photon emission is also performed within
the LCPI formalism \cite{LCPI1}. 
We reduce calculation of the photon emission rate
to solving a two dimensional Schr\"odinger equation.
For the pQCD model with static coupling constant and the thermal quark mass 
predicted by the HTL scheme our method 
is equivalent to the well known AMY formalism \cite{AMY1}.
We found that for the model of the QGP evolution that allows one
to obtain a reasonable description of jet quenching  
both the models for the photon emission 
underestimate considerably the photon spectrum
measured by PHENIX \cite{PHENIX_ph_PR}. 
For the phenomenological scenario with running $\alpha_s$ 
with a very small thermal quark mass
($m_q=50$ MeV) the contribution of the higher order collinear processes
summed with the LO $2\to 2$ processes can 
explain $\sim 50$\% 
of the experimental photon yield from PHENIX  \cite{PHENIX_ph_PR}
at $k_T\sim 2-3$ GeV.
Thus, we conclude that, 
for the picture of the QGP evolution and for the model of 
multiple parton scattering in the QGP consistent with data on jet quenching, 
the photon emission from the QGP stage alone is not enough to fit
the data on the photon production in Au+Au collisions
at $k_T\sim 2- 3$ GeV even for the scenario with a very small thermal quark 
mass.

\begin{acknowledgments} 	
This work has been supported by the RScF grant 16-12-10151.
\end{acknowledgments}

\end{document}